# Strain Gradient Plasticity-based Modeling of Hydrogen Environment Assisted Cracking

Emilio Martínez-Pañeda[1*], Christian F. Niordson[2], Richard P. Gangloff[3]

[1] Department of Construction and Manufacturing Engineering, University of Oviedo, Gijón 33203, Spain

[2] Department of Mechanical Engineering, Solid Mechanics, Technical University of Denmark, Kgs. Lyngby, DK-2800, Denmark

[3] Department of Materials Science and Engineering, University of Virginia, Charlottesville, VA 22904, USA

## ABSTRACT

Finite element analysis of stress about a blunt crack tip, emphasizing finite strain and phenomenological and mechanism-based strain gradient plasticity (SGP) formulations, is integrated with electrochemical assessment of occluded-crack tip hydrogen (H) solubility and two H-decohesion models to predict hydrogen environment assisted crack growth properties. SGP elevates crack tip geometrically necessary dislocation density and flow stress, with enhancement declining with increasing alloy strength. Elevated hydrostatic stress promotes high-trapped H concentration for crack tip damage; it is imperative to account for SGP in H cracking models. Predictions of the threshold stress intensity factor and H-diffusion limited Stage II crack growth rate agree with experimental data for a high strength austenitic Ni-Cu superalloy (Monel K-500) and two modern ultra-high strength martensitic steels (AerMet$^{TM}$100 and Ferrium$^{TM}$M54) stressed in 0.6M NaCl solution over a range of applied potential. For Monel K-500, $K_{TH}$ is accurately predicted versus cathodic potential using either classical or gradient-modified formulations; however, Stage II growth rate is best predicted by a SGP description of crack tip stress that justifies a critical distance of 1 µm. For steel, threshold and growth rate are best predicted using high-hydrostatic stress that exceeds 6 to 8 times alloy yield strength and extends 1 µm ahead of the crack tip. This stress is nearly achieved with a three-length phenomenological SGP formulation, but additional stress enhancement is needed, perhaps due to tip geometry or slip-microstructure.

**KEYWORDS:** Hydrogen embrittlement, multiscale simulations, electrochemistry, strain gradient plasticity, environment-assisted cracking

---

* Corresponding author. Tel.: +34985181967; fax: +34985182433.
*E-mail address*: mail@empaneda.com (E. Martínez-Pañeda)

# 1. INTRODUCTION

Multi-scale model predictions of material properties are important for alloy and process development, material life-cycle optimization, and component performance prognosis [1]. Interdisciplinary advances in deformation processing [2], fatigue [3], stress corrosion cracking (SCC) [4], and hydrogen embrittlement [5] illustrate this cutting-edge approach. Internal hydrogen and hydrogen environment assisted cracking (IHAC and HEAC, respectively) degrade high toughness alloys in fracture-critical aerospace, ship, energy, and ground transportation structures [6]. Moreover, hydrogen-stimulated damage is a primary mechanism for SCC of titanium, iron, nickel and aluminum-based alloys [7]. Models based on hydrogen-enhanced decohesion (HEDE) [8], interacting with hydrogen-enhanced localized plasticity (HELP) [9], predict trends in the subcritical crack growth rate properties of alloys stressed in environments that produce atomic hydrogen (H) via chemical and electrochemical reactions on crack tip surfaces [7,10]. However, improvements are required; local crack tip stress and dislocation configuration, as well as crack opening profile, are particularly important [11,12].

Building on elastic stress intensity factor (K) similitude for subcritical crack propagation [10], a diversity of IHAC and HEAC models [13-21] employ a crack tip stress field from classical continuum fracture mechanics [10,22], including finite-strain blunting [23], to predict growth threshold ($K_{TH}$) and rate (da/dt) properties. Alternative modeling is based on dislocation shielding of elastic crack tip stresses [24-27]. The difference between these two approaches centers on the magnitude and distribution of crack tip stresses, which define the location and severity of crack tip H-damage in the fracture process zone (FPZ). Continuum plasticity modeling shows that the maximum opening-direction tensile stress is 3-5 times alloy yield strength and located at 1-2 blunted crack tip opening displacements (of order 2 to 20 µm) ahead of the crack tip surface [23]. Dislocation-based models predict crack opening-direction stresses of 12-25 times yield strength and located much closer to the crack tip [24,25]. This difference is important because HEDE defines cracking as the balance between local tensile stress and H-concentration-reduced interface strength [8] (or reduced-total work of fracture [14,15]). Crack tip H concentration increases exponentially with rising hydrostatic stress [28,29], the crack tip stress gradient affects H diffusion [20,21], and dislocation density impacts the H flux via reversible-H trapping [21]. Next generation H-cracking models require an improved-quantitative description of the crack tip stress field between the extremes represented by classical continuum plasticity and dislocation shielding.

Extensive research has focused on the *smaller is harder* behavior of metals [30-34]. This size effect is attributed to geometrically necessary dislocations (GNDs), which accommodate lattice curvature due to non-uniform plastic deformation. Since classical plasticity lacks a material length, strain gradient plasticity (SGP) theories have been proposed to capture size effects. Isotropic SGP formulations are phenomenological (PSGP) [31] or mechanism-based (MSGP) [33,34]. These theories bridge the gap between length-independent continuum plasticity and discrete dislocation modeling by linking statistically stored and geometrically necessary dislocation densities to the mesoscale plastic strain/strain gradient and strain hardening. Since the plastic zone is small, with a large spatial gradient of high-strain deformation [23], it is imperative to account for GNDs in modeling crack tip stress and strain. Critically important for IHAC and HEAC, SGP modeling has consistently shown that increased GND density at the crack tip leads to: (a) higher local stresses, (b) a contraction in the breadth of the crack tip stress distribution, and (c) reduced blunting; each compared to classical plasticity [35-37]. SGP must be quantitatively implemented in material-damage models [38], as recognized for cleavage [39], interface fracture [40], layered-structure damage [41], ductile-microvoid fracture [42], fatigue [43], and H-enhanced cracking [7,44].

Recent SGP advances are relevant to finite element analysis (FEA) of crack tip stress and strain. PSGP theory with the full complement of three-gradient terms predicts high stresses that persist to a 10-fold larger distance ahead of the sharp crack tip compared to predictions from a single-length formulation [35]. However, this FEA was based on infinitesimal strain [31,35]. A large-strain FEA analysis of a blunting crack tip demonstrated that PSGP and MSGP formulations each predict elevated crack tip tensile stress and reduced crack tip opening compared to classical plasticity [36,37]. The distance over which this stress elevation persists is up to tens of micro-meters, sufficient to engulf the FPZ for HEAC [7], before merging with classical predictions at larger distances within the plastic zone. While classical plasticity predicts a stress maximum located at 1-2 blunted openings in front of the crack tip [22,23], large-strain SGP-enhanced stresses are highest at the smallest-FEA-modeled distance (100 nm) ahead of the tip, with no evidence of a stress maximum. Finally, SGP promotes stress elevation that depends on applied load, in sharp contrast to the $K_I$ independence of maximum stress predicted by classical plasticity [23]. The crack tip stress distribution is affected by both the SGP model used and value(s) of the material length(s). Uncertainties remain regarding: (a) the constitutive prescription that best captures increased GND density associated with a plastic strain gradient [32], and (b) the absolute values of material-dependent length(s) dependent on test method (e.g., nano-indentation) and SGP-model analysis of such measurements [35,37].

## 2. OBJECTIVE

The objective of this research is to implement and validate the coupling of a large-strain FEA-SGP analysis of crack tip stress with HEDE-mechanism-based models that predict HEAC propagation threshold and kinetics properties. Specific aims are to: (1) improve the basis for HEAC models using SGP inputs and insights, (2) predict H-cracking properties with fewer model parameters, (3) contribute insight into the role of GNDs ahead of a crack tip, and (4) experimentally assess the proper continuum-SGP formulation of crack tip stresses.

Model assessment is based on measurements of da/dt versus $K_I$ for HEAC in a Ni-Cu superalloy [45,46] and two ultra-high strength martensitic steels [47,48], each stressed in a chloride solution. Electrochemistry measurements and modeling yielded diffusible crack tip H concentration versus bold-surface applied potential ($E_{APP}$) [45,49], as well as trap-affected effective H diffusivity ($D_{H-EFF}$) for each alloy [50-52]. The $E_{APP}$ dependencies of $K_{TH}$ and the H-diffusion limited Stage II crack growth rate ($da/dt_{II}$) were originally modeled [45-48] using crack tip stress expected from blunt-crack [23] and dislocation shielding [24] analyses. This database and the HEDE-modeling approach are reanalyzed using crack tip stress distributions from new FEA that incorporates: (a) the finite strain framework for both PSGP and MSGP [37], and (b) specific alloy-dependent properties and load levels that create H cracking.

## 3. EXPERIMENTAL PROCEDURE

Three high strength alloys were modeled: (a) an austenitic Ni-Cu superalloy hardened by spherical γ' precipitates ($Ni_3(Al,Ti)$; 5 nm radius, 0.08-0.1 volume fraction, and 150000 to 190000 precipitates/µm$^3$ [53]), and (b) two martensitic ultra-high strength steels strengthened by needle-shaped carbide precipitates (($Cr,Mo)_2C$; 1 nm radius, 5-8 nm length, volume fraction of order 0.03, and about 150000 precipitates/µm$^3$ [51,54,55]). The heat treatment and microstructure of the superalloy, Monel K-500 (K500; Ni-28.6Cu-2.89Al-0.45Ti-0.166C by wt pct), are described elsewhere [45,50,53]: 0.2% offset yield strength ($\sigma_{YS}$) is 773 MPa, elastic modulus (E) is 183.9 GPa, and ultimate tensile strength ($\sigma_{UTS}$) is 1169 MPa from tensile testing; Ramberg-Osgood flow constants [56] from compression testing are n = 20, α = 0.39, E = 185.7 GPa and $\sigma_o = \sigma_{YSc}$ = 786 MPa; and plane strain fracture toughness ($K_{IC}$) is 200 to 340 MPa√m. The two similar quenched and aged block-martensitic alloy steels, AerMet$^{TM}$100 (AM100; Fe-13.4Co-11.1Ni-3.0Cr-1.2-Mo-0.23C by wt pct) and Ferrium$^{TM}$M54 (M54; Fe-7.0Co-10.1Ni-1.0Cr-2.1Mo-1.3-W-0.1V-0.30C by wt pct), are described elsewhere [47,48,51,54,55]. For AM100 and M54, respectively, $\sigma_{YS}$ is 1725 MPa and 1720 MPa and $\sigma_{UTS}$ is 1965 MPa and 2020 MPa from tensile testing; Ramberg-Osgood constants are n = 13 and 14, α = 1.0, E = 194 and 198 GPa, $\sigma_o = \sigma_{YSc}$ = 1985 MPa and 1951 MPa; and $K_{IC}$ is 130 MPa√m and 126 MPa√m.

The kinetics of HEAC were measured for K500, AM100, and M54 using precracked fracture mechanics specimens stressed under slow-rising $K_I$ while immersed in an aqueous solution of 0.6 M NaCl and as a function of $E_{APP}$, as detailed elsewhere [5,45-48]. The da/dt versus $K_I$ results for each alloy are typical of HEAC in high strength metals [7]. Two material properties characterize these data; specifically, the $K_{TH}$ for the onset of resolvable crack propagation under slow-rising $K_I$, which rapidly accelerates in Stage I then transitions in Stage II to K-independent growth at a plateau level (da/dt$_{II}$) due to chemical reaction or mass transport limitation [10]. The measured $E_{APP}$ dependencies of $K_{TH}$ and da/dt$_{II}$ (taken at a fixed $K_I$ of 40 to 50 MPa√m within Stage II) are used to assess the predictions of HEAC models that incorporate either MSGP or PSGP. All potentials are expressed with respect to the saturated calomel reference electrode, SCE.

## 4. MODELING PROCEDURE

### *4.1 Hydrogen assisted-cracking modeling*

$K_{TH}$ is modeled following the approach by Gerberich et al. that yielded [25]:

$$K_{TH} = \frac{1}{\beta'} \exp \frac{(k_{IG} - \alpha C_{H\sigma})^2}{\alpha'' \sigma_{YS}} \tag{1}$$

The β' and α'' constants, 0.2 (MPa√m)$^{-1}$ and 0.0002 MPa·m, respectively, are determined from numerical analysis of computer simulation results for an assumed configuration of dislocation shielding of the crack tip [24,57], and $C_{H\sigma}$ is defined below. The α (MPa√m per atom fraction H) is a weighting factor, which governs H-lowering of the Griffith toughness ($k_{IG}$, MPa√m), or the reversible work of fracture related to surface energy (γ$_S$) through $k_{IG}^2 = 2\gamma_s E/(1-\nu^2)$. The β' and α'' capture the impact of plasticity (plastic work of fracture) on this $\gamma_S$-based description. For the cases investigated, H diffusion from the crack tip into the FPZ likely governs the Stage II $da/dt_{II}$, modeled as [48,58,59]:

$$\left(\frac{da}{dt}\right)_{II} = \frac{4D_{H-EFF}}{x_{crit}} \left\{ \text{erf}^{-1}\left(1 - \frac{C_{H\sigma-crit}}{C_{H\sigma}}\right) \right\}^2 \quad (2)$$

where $x_{crit}$ is the critical distance ahead of the crack tip where H cracking nucleates leading to an increment of discontinuous crack advance. $C_{H\sigma-crit}$ is the critical concentration of H necessary for H decohesion at $x_{crit}$ and an inverse function of local tensile stress [8,60].

Consistent with the derivations of (1) and (2), $C_{H\sigma}$ must be the crack tip $\sigma_H$-enhanced concentration of mobile H in equilibrium with the crack tip overpotential for H production ($\eta_H$) and proximate to the interfacial-H crack path within the FPZ. Since $\sigma_H$ depends on distance ahead of the tip, $C_{H\sigma}$ varies with location, and is evaluated at $x_{crit}$ for use in (1) and (2). $C_{H\sigma}$ is derived as follows. The diffusible (or mobile) H concentration, $C_{H-Diff}$, is the sum of the normal-interstitial-lattice H ($C_L$) and reversibly trapped H ($C_{RT}$) for a single trap site, with $C_{RT}$ in local equilibrium with $C_L$ and described using Fermi-Dirac statics [28]. $C_L$ and $C_{RT}$ are of the same order for face-centered cubic K500 [50], but the reversible H concentration in body-centered martensitic steel is of orders of magnitude higher than $C_L$ [51]. $\sigma_H$ increases $C_L$ to $C_{L\sigma}$ due to lattice dilation [29], thus enhancing $C_{RT}$ in equilibrium with $C_{L\sigma}$ to yield $C_{H\sigma}$ [45]:

$$C_{H\sigma} = \left[C_L \frac{(1-C_{L\sigma})}{(1-C_L)} exp\left(\frac{\sigma_H V_H}{RT}\right)\right]\left[1 + \frac{(1-C_{RT})}{(1-C_L)} exp\left(\frac{E_B}{RT}\right)\right] \quad (3)$$

where $V_H$ is the partial molar volume of H in the metal lattice, $E_B$ is the binding energy of H to the dominant-reversible trap site adjacent to the crack path, T is temperature, and R is the gas constant. For H dissolved in the ultra-high strength steels and Ni-Cu superalloy in NaCl solution, $C_L$ and $C_{RT}$ are less than 0.001 atom fraction H, to justify that (1 – $C_L$) and (1 – $C_{RT}$) equal 1. $E_B$ for H in K500 and AM100 is 10 kJ/mol for H trapping at $Ni_3$(Al,Ti) or $(Cr,Mo)_2C$, respectively [50,51]. Therefore, the $E_B$ term in (3) is much greater than 1 and:

$$C_{H\sigma} = \left[(1 - C_{L\sigma})exp\left(\frac{\sigma_H V_H}{RT}\right)\right]\left[C_L exp\left(\frac{E_B}{RT}\right)\right] \quad (4)$$

The second-bracketed exponential term in (4) is $C_{RT}$, which essentially equals experimentally measurable $C_{H\text{-Diff}}$ and is elevated by $\sigma_H$ through the first-bracketed term. The $(1-C_{L\sigma})$ often equals 1 since $C_L$ is less than 0.001 wppm and $C_{L\sigma}$ is typically much less than 1.

Diffusible H concentration, unique to the occluded crack tip, must be determined to establish $C_{H\sigma}$ for $K_{TH}$ and $da/dt_{II}$ modelling in (1), (2) an (4). Measurements of artificial crevice pH and potential, coupled with a geometric model that scales crevice behavior to a tight crack, yielded the following relationship between $E_{APP}$, and crack tip H solubility ($C_{H,Diff}$) for K500 in aqueous chloride [45].

$$C_{H,Diff}(\text{wppm}) = -52.5 - 68.7 E_{APP}(V_{SCE}) \quad (5)$$

This result is relevant to HEAC in K500 with: (a) $E_{APP}$ less than -0.575 V, below the open circuit potential (OCP, about -0.225 V,) and (b) $10 < \xi < 60$ cm, where $\xi$ is the ratio of crack length squared to the average of crack mouth and blunt-tip openings. For AM100 in 0.6M NaCl at $E_{APP}$ below -0.750 V, the upper and lower bounds on crack tip H solubility are identical, and given by [49]:

$$C_{H,Diff}(\text{wppm}) = 19.125 E_{APP}^3 + 78.568 E_{APP}^2 + 80.026 E_{APP} + 24.560 \ (V_{SCE}) \quad (6)$$

for an HEAC-relevant $\xi$ of 15 to 20 cm (increasing $\xi$ from 10 to 1000 cm results in at most a 10% increase in $C_{H,Diff}$). For $E_{APP}$ between -0.750 V and -0.480 V, compared to the OCP of about -0.525 V, crack tip $C_{H,Diff}$ is less certain [49]. For example, $C_{H,Diff}$ increases from 1.7 to 2.8 wppm as $\xi$ rises from 10 to 1000 cm, with the latter typical of low $K_I$ (10-20 MPa√m) and restricted crack opening compared to classical blunting [23]. Moreover, H solubility is reduced to nearly 0 with increasing crack surface passivation [49]. Given these complications and limited data, for $E_{APP}$ above -0.750 V, crack tip H solubility for the two steels is given by (6) as the lower bound and the following upper bound [49]:

$$C_{H,Diff}(\text{wppm}) = -739.24E_{APP}^5 - 3121.1E_{APP}^4 - 5147.1E_{APP}^3 - 4099.2E_{APP}^2 - 1563.8E_{APP} - 225.77 \ (V_{SCE}) \quad (7)$$

The applied potential dependence of $K_{TH}$ is predicted by relating $E_{APP}$ to $C_{H\sigma}$ using (5) for K500, or (6) and (7) for the steels in (4) with the relevant $\sigma_H$ from SGP FEA, then fitting the single-unknown parameter, α, in (1) to $K_{TH}$ measured for any $E_{APP}$. A similar procedure is employed to predict the $E_{APP}$ dependence of $da/dt_{II}$ using (2), with measured $D_{H\text{-Eff}}$ [52] and independently determined $x_{crit}$ [7,59]. Critically, $da/dt_{II}$ is predicted without adjustable parameters since $C_{H\sigma}$ appears in (1) and (2). Equating (1) and (2) defines $C_{H\sigma-crit}$ as a function of α from $K_{TH}$ modeling, plus a single-measured $K_{TH}$ and $da/dt_{II}$ at any $E_{APP}$:

$$C_{H\sigma-crit} = \frac{1}{\alpha}\left(k_{IG} - \sqrt{\alpha''\sigma_{YS}\ln(K_{TH}\beta')}\right)\left[1 - \text{erf}\left(\sqrt{\frac{\left(\frac{da}{dt}\right)_{II} \cdot x_{crit}}{4D_H}}\right)\right] \quad (8)$$

$C_{H\sigma\text{-crit}}$ from (8) and $C_{H\sigma}$ from (4) must be evaluated at the same $K_I$; however, any value can be used since $C_{H\sigma\text{-crit}}/C_{H\sigma}$ is a constant independent of $\sigma_H$ and the associated $K_I$.

### *4.2 Strain gradient plasticity modeling*

PSGP [31] and MSGP models [34] were incorporated in an FEA of crack tip stress, as detailed elsewhere [36,37]. In the PSGP generalization of J2 flow theory [37], hardening due to the plastic strain gradient is incrementally captured through the generalized plastic strain rate ($\dot{E}_P$), formulated as a function of the conventional plastic strain rate ($\dot{\epsilon}^p$), three unique non-negative invariants ($I_i$) of $\dot{\epsilon}^p$, and three material lengths, $l_i$:

$$\dot{E}_P = \sqrt{\dot{\epsilon}^{p2} + l_1^2 I_1 + l_1^2 I_2 + l_1^2 I_3} \quad (9)$$

The MSGP formulation is based on the Taylor relationship between flow strength ($\sigma_{flow}$) and dislocation density, given by the sum of statistically stored ($\rho_S$) and geometrically necessary ($\rho_G$) dislocation densities [33,34]. The GND density is related to the effective plastic strain ($\varepsilon^p$) gradient ($\eta^p$) through the Nye-factor ($\bar{r}$) and Burger's vector ($b$):

$$\rho_G = \bar{r}\frac{\eta^p}{b} \quad (10)$$

These MSGP relationships predict flow strength as a function of $\varepsilon^p$, $\eta^p$, a single length parameter ($l$) and a reference stress ($\sigma_{ref}$) determined from the material flow rule [31,36]:

$$\sigma_{flow} = \sigma_{ref}\sqrt{f^2(\varepsilon^p) + l\eta^p} \qquad (11)$$

Since the Taylor dislocation model represents an average of dislocation activities, the MSGP theory is only applicable at a scale larger than the average dislocation spacing ($r \geq 100$ nm).

The material-dependent length is a single or multiple coefficient(s), calculated to fit experimental measurements of a size dependent property (e.g., hardness) using a specific SGP theory. Various micro-tests should be conducted to establish the $l_i$ parameter(s); however, this determination is outside the scope of the present work. The observed range of $l_i$ for metals is from 300 nm to 10 $\mu m$ (e.g., [30,61-63]). Reference lengths ($l = l_{ref}$ in MSGP and $l_1 = l_2 = l_3 = l_{ref}$ in PSGP) of 5 $\mu m$ for K500 and 7 $\mu m$ for AM100 are adopted. The former is based on micro-bending experiments with pure nickel [30], while the choice for AM100 rests on nano-indentation tests with a moderate strength steel [63]. A constant $l_{ref}$ is assumed in the PSGP model, as different weighting of individual length parameters has little influence in finite strain crack tip analyses [37]. The influence of length scale is addressed in the Discussion.

Crack tip stress analysis by boundary layer FEA, with PSGP and MSGP in the finite-strain framework, is detailed elsewhere [36,37]. $K_I$ quantifies the applied load, assuming plane strain and small-scale yielding. A refined mesh is used near the tip, where the length of the smallest element is 5 nm. The cracked body is discretized by 6,400 quadrilateral quadratic elements and the starting blunt-tip radius is $10^{-5}$-times the outer radius of the field [23].

## 5. RESULTS

### 5.1 Monel K-500

The crack tip hydrostatic stress distribution is computed for several applied $K_I$ in the range where HEAC occurred in K500. Figure 1 shows normalized $\sigma_H/\sigma_Y$ versus distance from the crack tip, $r$, for three cases: MSGP (with $l_{ref} = 5$ µm), PSGP (with $l_1 = l_2 = l_3 = l_{ref} = 5$ µm), and classical von Mises plasticity. All finite-strain, blunt-crack predictions agree beyond the location of maximum stress in the classical analysis, but significant differences arise closer to the crack tip. These findings are consistent with SGP results for a low strength-high work hardening alloy [37]. Specifically, for MSGP and PSGP compared to conventional plasticity: (1) crack tip stresses are substantially elevated, (2) a stress maximum is not evident, and (3) the stress distribution rises with increasing $K_I$. For the length(s) used, $\sigma_H$ from the 3-parameter PSGP model are higher than those predicted with MSGP. For each model, the maximum distance of 2.5 to 12 µm ahead of the crack tip where GNDs significantly influence the stress distribution suggests that SGP plays an important role in HEAC. Figure 2 shows MSGP-predicted GND density from (10) and the reduced crack tip profile in the opening (y) direction for each SGP formulation. The $\rho_S$ (Figure 2a) is determined from the uniaxial stress-strain curve [37], and the very high and localized GND density from SGP is apparent for each $K_I$ level. Crack tip opening (Figure 2b) is reduced by hardening from this high $\rho_G$.

For HEAC modeling, crack tip $\sigma_H$ is averaged over two distances, 0.1 µm < $r$ < 1 µm and 0.1 µm < $r$ < 2 µm, as justified in the Discussion, and values are given in Table 1 including results for a low strength alloy [37]. $K_{TH}$ from (1) is predicted versus $E_{APP}$ for the PSGP and MSGP-model values of $\sigma_H/\sigma_Y$ (the average of the 1 µm and 2 µm intervals of $r$, Table 1) using $C_{H\sigma}$ from (4) and (5). Model results in Figure 3a are compared to experimental data for K500 in 0.6M NaCl solution [45,46]. The 3-replicate measurements of $K_{TH}$ at $E_{APP}$ of -1.000 $V_{SCE}$ are used to determine α, which equals 6.36 MPa√m(at frac H)$^{-1}$ for PSGP-based $\sigma_H$ (8.1$\sigma_Y$) and 37.59 MPa√m(at frac H)$^{-1}$ for MSGP $\sigma_H$ (4.7$\sigma_Y$). The remaining constants in (1) were justified, including $k_{IG}$ of 0.880 MPa√m from $\gamma_S$ for Ni [45]. Since $C_L$ is 1 to 50 wppm for Monel K-500 in NaCl solution [45], (1 – $C_{L\sigma}$) is essentially 1.0 in (4). The PSGP and MSGP-based predictions of $K_{TH}$ similarly agree with measured values over a range of $E_{APP}$; only α rises as crack tip $σ_H$ falls. Each α from the Figure 3a fit is used to calculate a $C_{Hσ\text{-crit}}$ through (8) with $K_{TH}$ and $da/dt_{II}$ measured at $E_{APP}$ of -1.000 $V_{SCE}$. The $da/dt_{II}$ is then calculated from (2) and the results are given in Figure 3b. The PSGP and MSGP predictions of $da/dt_{II}$ are essentially identical, and agree with measured $da/dt_{II}$ at a single $K_I$ of 50 MPa√m [45,46]. [2]

Table 1. Large strain FEA predictions of $σ_H/σ_Y$, at $r$ = 1 or 2 μm ahead of the blunted crack tip for conventional plasticity, and averaged between the blunted crack tip and $r$ = 1 or 2 μm for two SGP formulations with $l_{ref}$ = 5 μm for Monel K-500 and $l_{ref}$ = 7 μm for AerMet™100.

| $σ_H/σ_Y$ | $K_I$ (MPa√m) | Classical ($r$ = 1, 2 μm) | MSGP ($r$ = 1, 2 μm) | PSGP ($r$ = 1, 2 μm) | Elastic Singular ($r$ = 0.25, 1 μm) |
|---|---|---|---|---|---|
| **AerMet™100** (Figure 4) | 10 | 1.8, 1.7 | 2.2, 2.0 | 2.8, 2.5 | 4.1, 2.1 |
| | 20 | 1.4, 1.6 | 3.4, 3.1 | 4.6, 4.1 | 8.2, 4.2 |
| | 40 | 0.8, 1.1 | 5.5, 5.1 | 7.6, 6.8 | 16.4, 8.4 |
| | 80 | 0.5, 0.8 | 8.6, 8.1 | 14.0, 13.2 | 32.8, 16.8 |
| **Monel K-500** (Figure 1) | 17.3 | 1.5, 1.8 | 4.8, 4.6 | 8.6, 7.7 | 15.9, 7.9 |
| | 50 | 1.0, 1.1 | 7.1, 6.7 | 16.8, 16.5 | 45.9, 22.8 |
| **Low Strength** [37] | 22.4 | 2.8, 3.6 | 10.4, 9.1 | 21.0, 16.0 | 39.8, 19.9 |

*5.2 AerMet™ 100 and Ferrium™ M54*

The crack tip hydrostatic stress distribution is computed for several $K_I$ relevant to HEAC of AM100 and M54. Figure 4 shows $σ_H/σ_Y$ versus $r$ for MSGP ($l_{ref}$ = 7 μm), PSGP ($l_{ref} = l_1 = l_2 = l_3$ = 7 μm), and classical plasticity. Stresses are given in Table 1, and show the same behavior as K500 (Figure 1) and a low strength alloy [37].

$K_{TH}$ versus $E_{APP}$ is predicted from (1) and (4) using crack tip H solubility from either the upper bound given by (6) and (7) or the lower-bound in (6); the results are presented in

---

[2] Filled points in Figure 3 represent 100% intergranular HEAC, while open points with upward arrows show those $E_{APP}$ that did not produce intergranular HEAC for the highest-applied $K_I$ [46]. The two points at $E_{APP}$ (■) of -0.900 and -0.800 V were associated with intergranular HEAC attributed to specimen-to-specimen variability in grain boundary-S segregation. This behavior was captured by higher α, lower $k_{IG}$, and lower $C_{Hσ\text{-crit}}$ than used for the majority of $K_{TH}$ and $da/dt_{II}$ measurements in Figure 3 [46]. These parameter changes are consistent with grain boundary weakening due to S interaction with H.

Figures 5 and 6. Parts (a) and (b) of each figure show the PSGP and MSGP results, respectively. The three levels of averaged $\sigma_H/\sigma_Y$ (Table 1) correspond to $K_I$ of 10 MPa√m, 20 MPa√m, and 40 MPa√m. The $k_{IG}$ is 1.145 MPa√m for each steel, and the α" and β' are identical to those used for K500 [46] and steel [25]. Griffith toughness was estimated based on maximum modeled $\gamma_S$ for a {100} surface of Fe (3.09 J/m$^2$ [64]) and Poison's ratio of 0.29. This $k_{IG}$ yielded a H-free $K_{IC}$ of 224 MPa√m through (1), which is reasonably higher than the intervening microvoid based $K_{IC}$ (130 MPa√m). However, the precise Griffith toughness for a HEDE-sensitive martensite block or packet interface in AM100 and M54 is not known [47]. Each SGP prediction is given by a solid plus dashed curve, and compared to experimental measurements of $K_{TH}$ [49,50].[3] For each case examined, an average α is calculated using the six experimental values of $K_{TH}$ at $E_{APP}$ of -0.900 V and lower. This regime was selected because H solubility is well known through (6), HEAC is severe (measured $K_{TH}$ varied between 9 MPa√m and 14 MPa√m with an average of 10.5 MPa√m), HEAC is reproducible (3 replicated values are essentially equal for M54 at $E_{APP}$ of -1.000 V), and HEAC is fully transgranular associated with martensite interface decohesion [47]. Average-calculated α values are given in Figures 5 and 6. The dashed curves show the regime of $E_{APP}$ where the $K_{TH}$ model from (1) is expected to under-predict true $K_{TH}$ for HEAC, as justified in the Discussion.

The $E_{APP}$ dependence of $da/dt_{II}$ is predicted without any adjustable constants using independently established $D_{H-EFF}$ [52] and $x_{crit}$ [59]; results are shown for upper bound (Figure 7) and lower bound (Figure 8) $C_{H,Diff}$. PSGP ($\sigma_H = 7.2\sigma_Y$, solid line) and MSGP ($\sigma_H = 5.3\sigma_Y$, dashed line) predictions are shown in each plot, and compared to $da/dt_{II}$ measured at a $K_I$ of 40 MPa√m [47,48]. Each $C_{H\sigma-crit}$ is calculated through (8), using the appropriate α from the Figures 5 and 6 fits at $K_I$ of 40 MPa√m, coupled with the average $K_{TH}$ and average da/dt$_{II}$ measured at E$_{APP}$ of -1.000 V. Downward arrows represent experiments where K$_{TH}$ exceeded 40 MPa√m, and HEAC was not resolved; all other data are associated with transgranular HEAC [47,48]. The predictions of the SGP-HEAC model in Figures 5 through 8 effectively capture the complex dependencies of K$_{TH}$ and da/dt$_{II}$ over a wide range of E$_{APP}$.

---

[3] The largest $C_{H-Diff}$ is 6 wppm, and $C_L$ is about 0.06 wppm, at the most cathodic $E_{APP}$ examined. As such, $C_{L\sigma}$ is 0.01 atom fraction H for the highest $\sigma_H/\sigma_Y$ of 7.2 and the calculations in Figure 5 equate (1-$C_{L\sigma}$) in (4) to 1.

## 6. DISCUSSION

### *6.1 SGP Impact on Hydrogen Cracking*

Strain gradient plasticity (SGP) enhanced large-strain finite element analysis (FEA) results reveal a profound influence of geometrically necessary dislocation (GND) density on crack tip mechanics for technologically important alloys. Simulation results in Figures 1, 2 and 4 establish the following effects of mechanism-based (MSGP) and phenomenological (PSGP) strain gradient plasticity compared to classical plasticity analysis of a blunt crack tip.

- Crack tip stresses are substantially elevated, and crack opening is reduced, due to hardening from high-GND density. This reduced CTOD is strictly a continuum mechanics effect, which is not related to H-plasticity interaction that could impact the local slip mode, hardening/softening, or crack path through the microstructure [9].

- σ$_H$ levels from the 3-parameter PSGP model are substantially higher than those predicted by the MSGP formulation.

- The maximum in tensile stress with increasing distance is shifted to within 100 nm or less from the blunted crack tip by SGP hardening.

- The crack tip stress distribution from SGP rises and broadens with increasing K$_I$.

- The magnitude of SGP-elevated $\sigma_H/\sigma_Y$ decreases with increasing alloy strength and the maximum crack tip $\sigma_H$ is essentially constant (6300 MPa or ~0.035E)[4].

- GND density and $\sigma_H$ are elevated over 1 to 20 µm ahead of the crack tip, suggesting that SGP impacts hydrogen (H) cracking in the fracture process zone (FPZ).

It is imperative to account for the strain gradient in modeling of hydrogen environment (HEAC) and internal hydrogen (IHAC) assisted cracking over a wide range of alloy strengths.

*6.2 Fracture Process Zone Definition*

A critical distance, $x_{crit}$, from the crack tip surface to FPZ sites of H damage formation, is required to define crack tip $\sigma_H$ to calculate $C_{H\sigma}$ through (4) and $da/dt_{II}$ in (2). Classical plasticity equates this distance to the location of maximum stress [13-23], evident in Figures 1 and 4. This classical $x_{crit}$ is 6 to 13 µm for K500 at $K_I$ of 25 to 45 MPa√m and 5 to 10 µm for AM100 at $K_I$ of 30 to 50 MPa√m. In contrast empirical analysis suggests that $x_{crit}$ is 1 µm for alloys of different strengths and wide ranging $K_I$ [59]. A micro-meter-scale critical distance is consistent with the SGP predictions in Figures 1 and 4.

The SGP results suggest that $x_{crit}$ is the location of the highest probability of H-assisted crack formation, governed by interaction of decreasing $\sigma_H$ (and decreasing $C_{H\sigma}$) with the increasing number of defect-based initiation sites within the FPZ; each with increasing *r*. The details of H-crack formation are not sufficiently defined to quantify $x_{crit}$, following the approach used to model cleavage [65]. Electron microscopy suggests that hydrogen-enhanced localized plasticity (HELP) concentrates stress to promote interface hydrogen-enhanced decohesion (HEDE) [9]. Speculatively, the number of crack formation sites scales with $\rho_G$ and interacts with $C_{H\sigma}$ to establish $x_{crit}$. For K500, GND density from MSGP is above $\rho_S$ for *r* up to 0.5 µm at $K_I$ of 15 MPa√m and 2.2 µm at 45 MPa√m (Figure 2). Similar behavior is suggested for AM100, since $\sigma_H$ is elevated by MSGP for $r$ of up to 1 to 6 µm for HEAC relevant $K_I$ of 10 MPa√m to 40 MPa√m (Figure 4).

---

[4] Regression analysis of the PSGP simulation results (at $K_I$=20 MPa√m, averaged over the two intervals of *r* for the alloys in Table 1) yields $\sigma_H/\sigma_{YS}$ =6300/$\sigma_Y$ (in MPa).

Reversible H trapping at precipitate-matrix interfaces is extensive within an µm-scale FPZ for both alloys studied. The small size (1 to 5 nm) and large number density of (Ni$_3$(Al,Ti)) spheres and (Cr,Mo)$_2$C needles results in a mean-free path between precipitate surfaces of 25-40 nm for the steels and 60-75 for K500. Thermal desorption analysis affirmed that up to $1.5 \times 10^8$ H atoms (32 wppm) are trapped by monolayer coverage on all (Cr,Mo)$_2$C - surface sites in 1 µm$^3$ of AM100 for a single H overpotential [51]. Additionally, $C_L$ is a significant fraction of $C_{H,Diff}$ for the fcc superalloy and interstitial jump distance is of order 1 nm [50]. It follows that (1) through (4) provide a physically reasonable description of HEAC for the alloys considered.

Measurements that affirm $x_{crit}$ are not widely available. Micro-meter spaced makings attributed to H cracking were convincingly demonstrated for oriented-single crystal Fe-Si, and a martensitic steel [25,66]. These results not-with-standing, the small $x_{crit}$ challenges measurements using SEM, acoustic emission, electrical potential, or electrochemical current. Markings associated with $x_{crit}$ were not observed by SEM analysis of K500, AM100, or M54 for the cases modeled [45,47,48]. The martensitic microstructure of these steels, which constitutes the transgranular HEAC path [47,48], obscures crack advance markings, and a blunting-based feature may not occur in the short time (1 to 1000 s from Figures 7 and 8) between crack advance in a creep resistant alloy. For lower strength K500, intergranular HEAC features are more likely to show markings, but these can be either crack wake slip steps (not relevant to crack advance or due to discontinuous advance over $x_{crit}$. Work is required to characterize the site of the crack tip FPZ.

### *6.3 Crack Growth Rate Modeling*

To model HEAC, $x_{crit}$ was taken as 1.0 µm as a proxy for statistical analysis, and the average of the two stress levels in Table 1 was used for each SGP model, alloy, and $K_I$. Figures 7 and 8 show that measured and model predicted $da/dt_{II}$ agree precisely for AM100 at the most cathodic $E_{APP}$ examined. Here, for the high-PSGP stress level, severe HEAC is diffusion controlled and the combination of independently measured $D_{H-EFF}$ and $x_{crit}$ of 1.0 µm predicts measured $da/dt_{II}$ through (2). Reasonable agreement is observed for K500 at the most cathodic $E_{APP}$ below -1.000 V (Figure 3b); however, $x_{crit}$ would have to equal 0.35 µm for precise-model agreement with the single-highest $da/dt_{II}$. SGP modeling justifies an $x_{crit}$ of order 1 µm for HEAC, at least within the accuracy and relevance of measured $D_{H-EFF}$ [67,68].

The distributions of crack tip $\sigma_H$ and $\rho_G$ from SGP-FEA simulation can improve the accuracy of H diffusion models pertinent to HEAC and IHAC. The $da/dt_{II}$ model in (2) does not include the effects of crack tip stress on H flux and dislocation trapping of H on $D_{H-EFF}$ (typically from a stress-free H permeation experiment and approximate trapping analysis [67]). Sophisticated models address such complications [20,21,68]; however, these center on blunt-crack $\sigma_H$ and $\rho_S$ associated with plastic strain from classical plasticity [23]. In these models, the maximum crack tip $\sigma_H$ provides a positive stress gradient ahead of the crack tip, which increases the flux of H from the tip surface to $x_{crit}$ [19-21,58]. However, $\sigma_H$ monotonically declines with increasing $r$ due to SGP, at least for distances greater than 100 nm (Figures 1 and 4); $d\sigma_H/dr$ is mildly negative for MSGP and more strongly so for PSGP. The SGP-stress gradient retards H diffusion to $x_{crit}$. Second, the GND distribution due to SGP (Figure 2a) provides dislocation sites for reversible-H trapping that reduce $D_{H-EFF}$. Provided the binding energy of H to GND structure is known, equilibrium trapping theory can estimate the effect of dislocation density on the H diffusivity distribution relevant to the FPZ [28,67].

SGP modeling (Table 1) establishes that crack tip tensile stress rises with increasing $K_I$, which appears to be at odds with $K_I$ independent $da/dt$ in Stage II [7,10]. For example, $\sigma_H$ rises from $7.2\sigma_{YS}$ to $16.7\sigma_{YS}$ as $K_I$ increases from 40 MPa√m to 80 MPa√m for AM100 (PSGP, Table 1), but $da/dt$ is constant [47,48]. The H-diffusion model in (2) shows that

da/dt$_{II}$ depends on $C_{H\sigma\text{-crit}}/C_{H\sigma}$; critically, this ratio is independent of $K_I$ since each concentration is amplified by the same exponential dependence on $\sigma_H$ through (3) and (4).[5] Any $K_I$ can be used; however, a lower $K_I$ somewhat above $K_{TH}$ reduces $C_{H\sigma}$. When $C_{H\sigma}$ is large (~0.5 to 1.0 atom fraction H), stress due to lattice expansion from H in interstitial sites offsets the lattice dilating impact of $\sigma_H$ [24,69]. This issue is important for ultra-high strength steel, high $K_I$, and PSGP models (Table 1) where $\sigma_H/\sigma_{YS}$ above 9 results in unrealistic values of $C_{H\sigma}$ exceeding 1.0 atom fraction.

### *6.4 SGP-HEAC Model Validation*

The results of the present investigation affirm the integration of cutting edge SGP-FEA formulations with crack electrochemistry and two HEAC models to predict material-environment properties, specifically $K_{TH}$ and da/dt$_{II}$ as a function of environmental H activity. Models with a single calibration constant are validated over a broad range of applied polarization using precise experimental measurements of these HEAC properties. Excellent agreement is reported for a Ni-Cu superalloy with cathodic $E_{APP}$. The comparison for two ultra-high strength steels is good, but hindered by crack mechanics and electrochemical uncertainties.

*6.4.1 Monel K-500* The SGP-based predictions of $K_{TH}$ and da/dt$_{II}$ versus $E_{APP}$ quantitatively agree with experimental measurements for a single lot of K500 stressed under slow-rising $K_I$ in 0.6M NaCl solution with cathodic polarization. Occluded-crack electrochemistry was previously detailed [45,50], as was specimen variability due to grain boundary S segregation (Footnote 2) [45,46]. The one-length-parameter MSGP and three-term PSGP models of crack tip $\sigma_H/\sigma_{YS}$ similarly predict the applied potential dependence of $K_{TH}$ that agrees with experimental measurements over a range of cathodic $E_{APP}$ (Figure 3a).

---

[5] This ratio is determined by calculation of $C_{H\sigma}$ at any $\sigma_H$ (or $K_I$), followed by determination of α in (1) and $C_{H\sigma\text{-crit}}$ through (8) using the same $\sigma_H$. As a check for K500 with $C_{H\sigma}$ calculated from (5) at $E_{APP}$ = -1.000 V, $C_{H\sigma}/C_{H\sigma\text{-crit}}$ = 3.25 for $\sigma_H/\sigma_{YS}$ of 8.15 and $C_{H\sigma}/C_{H\sigma\text{-crit}}$ = 3.01 for $\sigma_H/\sigma_{YS}$ of 4.70. This 10% difference in $C_{H\sigma}/C_{H\sigma\text{-crit}}$ is not significant.

Moreover, $C_{H\sigma\text{-crit}}$ calculated from $K_{TH}$-calibrated α predicts the $E_{APP}$ dependence of $da/dt_{II}$ that agrees equally well with experimental measurements for both MSGP and PSGP (Figure 3b). Since only α was calibrated at a single-low $E_{APP}$ (-1.000 V) to model $K_{TH}$, with all other parameters in (2) (α", β', and $k_{IG}$) justified [46], and since no adjustable parameters were used to predict $da/dt_{II}$, the models represented by (1) and (2) are validated and consistent. The impact is clear; the wide-range dependence of HEAC properties on cathodic polarization is predicted with α calibrated at a single $E_{APP}$. This prediction includes an accurate value of the technologically critical potential, above which HEAC is eliminated.

Considering classical plasticity for $K_{TH}$ of 17.3 MPa√m, $\sigma_H/\sigma_{YS}$ is 1.5 at 1 μm ahead of the crack tip and 2.6 at the location ($r$ = 3 μm) of the maximum stress (Figure 1). Predictions of $K_{TH}$ versus $E_{APP}$ using either of these $\sigma_H$ levels agree with experimental values with α of 209.5 MPa√m(atom frac H)$^{-1}$ for $\sigma_H/\sigma_{YS}$ = 1.5 giving $C_{H\sigma\text{-crit}}$ of 12.3 wppm through (8) or α of 116.0 MPa√m(atom frac H)$^{-1}$ for $\sigma_H/\sigma_{YS}$ = 2.6 giving $C_{H\sigma\text{-crit}}$ of 22.2 wppm. The $K_{TH}$ versus $E_{APP}$ agreement form these classical plasticity predictions is essentially identical to the SGP-based results in Figure 3a. However, the stress maximum in the classical model suggests that $x_{crit}$ is 3 to 14 μm, for $K_I$ between 15 and 45 MPa√m, rather than 1 μm justified by SGP. As such, classical plasticity-based predictions of $da/dt_{II}$ are reduced by 3-fold to 14-fold at any $E_{APP}$ compared to the SGP curves in Figure 3b where $x_{crit}$ = 1 μm. While $K_{TH}$ modeling does not distinguish the most accurate crack tip stress field, the SGP models provide more accurate predictions of $da/dt_{II}$ compared to classical plasticity. This comparison supports the relevance of crack tip stress elevation due to GNDs.

*6.4.2 AerMet$^{TM}$100 and Ferrium$^{TM}$M54* SGP-HEAC model predictions of $K_{TH}$ and $da/dt_{II}$ versus $E_{APP}$ agree with measurements for AM100 and M54 stressed under slow-rising $K_I$ in 0.6M NaCl solution, as shown in Figures 5 through 8. First, absolute values of $K_{TH}$ at potentials above -0.600 V are accurately predicted using the single α calibrated at low $E_{APP}$ (Figure 5). In each regime transgranular HEAC is severe. Agreement is quantitatively strong for the highest level of crack tip stress from the PSGP simulation in Figure 5a. Second, the window of $E_{APP}$ between -0.600 and -0.800 V, where $K_{TH}$ rises sharply and $da/dt_{II}$ falls toward zero, is captured, as governed by the minimum in $C_{H\text{-Diff}}$ versus $E_{APP}$ given by (6) and (7). Third, reasonable predictions of $da/dt_{II}$ without adjustable parameters, using $C_{H\sigma\text{-crit}}$ calculated from α, demonstrates the consistency of the HEAC models given by (1) and (2).

Model assessment is demanding for steels given the change in occluded crack chemistry, which accompanies transition from cathodic to anodic polarization through the open circuit potential (OCP) of about -0.525 V. Crack tip $C_{H,Diff}$ is uncertain for $E_{APP}$ above about -0.750 mV owing to limited crack chemistry measurements and the effect of surface passivation [49]. It is only possible to bound $C_{H,Diff}$ using (6) and (7), leading to the upper and lower bound predictions of $K_{TH}$ (Figure 5 and 6) and $da/dt_{II}$ (Figures 7 and 8). The best prediction of the $E_{APP}$ dependence of these HEAC properties likely resides between these bounds. Second, the dashed parts of the predicted curves in Figures 5 and 6 show the regime of $E_{APP}$ where $C_{H\text{-Diff}}$ is less than 0.8 wppm and should promote mixed transgranular H-cracking and ductile microvoid fracture [70]. These dashed lines should under-predict measured $K_{TH}$ since the HEAC model in (1) does not capture the added cracking resistance associated with ductile growth. Third, $K_{TH}$ and low $da/dt$ are difficult to measure when plasticity at higher $K_I$ gives a false indication of low-rate crack extension from electrical potential measurement [47]. The variability of measured $K_{TH}$ for -0.800 V < $E_{APP}$ < -0.625 V is due in part to this limitation. Finally, surface reaction may interact with H diffusion for $E_{APP}$ below about -0.750 V [71]. The $da/dt_{II}$ from the H diffusion model in (2) is an upper bound when surface reaction is slow.

With these considerations, Figures 5 through 8 establish that the best agreement between measured and predicted $K_{TH}$ and $da/dt_{II}$ is achieved over a wide range of $E_{APP}$ for PSGP-based $\sigma_H/\sigma_Y$ of 7.2. These figures suggest that $\sigma_H$ as low as 6.0$\sigma_Y$ provides similar-good predictions. However, lower crack tip stress levels (2.1 < $\sigma_H/\sigma_Y$ < 5.3) provide poor agreement between measured and predicted HEAC properties for either the upper or lower bound H solubilities. For this high $\sigma_H$ regime, the bounds of crack tip H solubility in (6) and (7) are affirmed, as is evident by comparison of the solid line predictions of $da/dt_{II}$ versus $E_{APP}$ above -0.800 V in Figures 7 and 8 (speculatively, growth rates for $E_{APP}$ below -0.850 V are lower than the H-diffusion model prediction due to surface reaction rate limitation [71]). The $K_{TH}$ versus $E_{APP}$ predictions are mixed. Upper bound H solubility provides the best-absolute agreement in $K_{TH}$ for $E_{APP}$ above about -0.600 V and below -0.700 V (Figure 5a), but the lower bound $C_{H,Diff}$ relationship (Figure 6a) better captures the range of $E_{APP}$ (-0.770 to -0.585 V) where the dashed line defines the lower bound on the variability in $K_{TH}$ explained by plasticity-microvoid cracking and hindered crack growth resolution. It is likely that specimen to specimen differences are amplified for $E_{APP}$ above about -0.800 V due to the sensitivity of crack tip H production and uptake to small changes in: (a) crack surface passivity (reduced by acidification and Cl$^-$ intrusion), and/or (b) the magnitude of crack tip potential reduction below $E_{APP}$ (due to increased crack tip occlusion from corrosion product deposition [49]).

Considering classical plasticity analysis, the very low $\sigma_H/\sigma_Y$ at $x_{crit}$ of 1.0 to 2.0 µm (0.8 to 1.8, Table 1), or at the location of maximum stress (r = 1.4 to 12 µm, Figure 4), provide poor predictions of $K_{TH}$ and $da/dt_{II}$ versus $E_{APP}$. Such predictions are similar to those from the lower $\sigma_H/\sigma_Y$ SGP models in Figures 5 through 8. Moreover, $x_{crit}$ defined at the $\sigma_H$ maximum, predicts $da/dt_{II}$ that are substantially below measured values. Overall, the comparisons in Figures 5 through 8 establish the necessity for high crack tip $\sigma_H$, equal or above $6\sigma_Y$, in order to predict the wide-range $E_{APP}$ dependencies of $K_{TH}$ and $da/dt_{II}$ for steel. This result justifies both crack tip SGP and the relevance of the three-parameter PSGP formulation. However, this finding is problematic for $K_{TH}$ modeling because Table 1 shows that $\sigma_H/\sigma_Y$ above 6 is only predicted by the large strain FEA-PSGP analysis for $K_I$ of 35 to 40

MPa√m. It is necessary to identify the cause of high crack tip stresses for $K_I$ below 20 MPa√m.

It is difficult to justify very high crack tip stresses for ultra-high strength steel using the blunt crack PSGP approach *per se*. First, it is unlikely that the requirement for high crack tip stresses will be relaxed by changes in other aspects of the HEAC models. The parameters in the $K_{TH}$ model (α", β', and $k_{IG}$ in (1)) and $da/dt_{II}$ model ($D_{H-EFF}$ in (2)) were independently justified [45,46,50,52] and are consistent with the original analysis by Gerberich and coworkers [24,25,57]. Second, $l_i$ is a primary uncertainty in the PSGP and MSGP models, and has not been reported for ultra-high strength steel with a fine-scale martensitic structure and high $ρ_S$ ($10^{16}$ m$^{-2}$ [72]) without strain hardening. As such, an SGP-FEA sensitivity study was conducted for a single $K_I$ (20 MPa√m). In both SGP formulations, $σ_H/σ_Y$ (at $r$ < 2-5 μm) rises as $l_{ref}$ increases from 1 to 15 μm. For example, at $r$ = 1 μm, $σ_H/σ_{YS}$ rises from 2.1 to 3.5 for MSGP and from 1.8 to 3.8 for PSGP, as $l_i$ increases from 1 μm to 15 μm. These $σ_H$ elevations do not achieve 6 to 7-times $σ_Y$, extending over $r$ of 1-2 μm, as necessary to accurately predict $E_{APP}$ dependent $K_{TH}$ and $da/dt_{II}$ for Stage II $K_I$ below about 30 MPa√m. There is no indication that alternate values of $l_1$, $l_2$, and $l_3$ yield such high crack tip stresses.

Other approaches predict high crack tip stresses, but only over distances that are small compared to an $x_{crit}$ of 1 μm. As an upper bound, $σ_H$ from the singular terms of the plane strain elastic crack tip stress distribution is shown in Table 1. For the high strength steel, this stress exceeds $7σ_Y$ at $r$ = 1 μm, but only for $K_I$ above 33 MPa√m; even singular-elastic stresses are not sufficient. Dislocation free zone (DFZ) models show that the net crack tip stress field is reduced below the singular-elastic field [26,27]. The model represented by (1) is based on a DFZ approach, with the elastic crack tip stress field shielded by a pile-up of dislocations on a single slip plane coupled with a super-dislocation to capture the "far field" plastic zone [25]. Very high crack tip $σ_H/σ_Y$ is predicted, but only over $r$ less than 100 nm

[24].

Enhancements to the continuum large-strain elastic-plastic SGP-FEA analysis could explain very high crack tip stresses extending of order ~1 μm ahead of the crack tip. The PSGP and MSGP stress fields (Figures 1 and 4) were calculated for a smoothly blunting crack (e.g., Figure 2a) [23]. SGP hardening is likely to be elevated for a geometrically "sharp" or irregular crack tip with reduced relaxation of the singularity. A tip that blunts to form a sharp corner could promote locally high stresses not relaxed by regular-geometric blunting [73]. Tip shape may be controlled by microstructural enforcement of the HEAC path, typically localized along austenite grain boundaries in Ni-superalloys and lath-martensite interfaces in modern steels. Slip morphology, influenced by HELP [9], could impact crack tip shape. *In situ* loading and SEM stereo imaging of transgranular fatigue crack and intergranular HEAC tips demonstrated much less blunting for the latter [74]. Alternately, microstructure-scale stresses can be elevated by slip morphology, dislocation substructure, and grain-elastic anisotropy [19]. Research must establish HEAC tip shape evolution over a range of $K_I$, and integrate local strain hardening due to SGP-GNDs with microstructure-scale stresses, all captured in a finite-strain crack tip FEA.

## 7.0  CONCLUSIONS

Large strain finite element analysis of crack tip stress, augmented by phenomenological and mechanism-based strain gradient plasticity formulations for a blunt crack, is integrated with electrochemical assessment of occluded-crack tip H solubility and H-decohesion based damage models to predict hydrogen assisted crack growth properties. Predictions agree with a robust data base for a high strength Ni superalloy and two modern ultra-high strength martensitic steels stressed in an aqueous H-producing environment. Conclusions are as follows.

- Large-strain FEA models establish a profound influence of SGP on crack tip stress and strain; GND density increases, crack tip stresses are elevated but do not exhibit a near-tip maximum, and crack opening is reduced compared to classical blunt-crack plasticity.

- The impact of SGP decreases with increasing alloy strength, but in all cases hydrostatic stress enhancement leads to locally high crack tip H concentration to enable damage; it is imperative to account for SGP hardening in modeling of H cracking.

- Integrated SGP, occluded-crack electrochemistry, and HEAC models effectively predict the dependencies of threshold stress intensity and H-diffusion limited Stage II crack growth rate on applied electrode potential for Monel K-500 and ultra-high strength steel (AerMet$^{TM}$100 and Ferrium$^{TM}$M54) in NaCl solution with a single calibration constant.

- For Monel with cathodic polarization, $K_{TH}$ is accurately predicted using classical and SGP formulations of stress; however, Stage II crack growth rate is best predicted by the SGP descriptions that justify a critical distance of 1 µm due to crack tip stress elevation from GND hardening.

- For AerMet$^{TM}$100 and Ferrium$^{TM}$M54, measured and modeled $K_{TH}$ and $da/dt_{II}$ quantitatively agree for cathodic and anodic potentials, within the bounds of somewhat uncertain crack tip H solubility, but only for crack tip $\sigma_H/\sigma_Y$ of 6 to 8, which justifies SGP hardening and the relevance of a three-length PSGP model.

- Such high levels of crack tip $\sigma_H/\sigma_Y$, extending 1 µm beyond the crack tip, are not sufficiently predicted by PSGP simulation for low $K_I$ typical of $K_{TH}$ for the steels. The necessary-high stress is speculatively attributed to SGP interacting with crack tip geometry and/or HELP-sensitive microstructure-scale stresses.

**ACKNOWLEDGEMENTS**

E. Martínez-Pañeda acknowledges financial support from the Ministry of Science and Innovation of Spain (grant MAT2011-29796-C03-03), and from the University of Oviedo (UNOV-13-PF). C.F. Niordson acknowledges support from the Danish Council for Independent Research under the research career program Sapere Aude in the project "Higher Order Theories in Solid Mechanics". R.P. Gangloff acknowledges support from the Faculty Affiliate programs of the Alcoa Technical Center and the Northrup Grumman Corporation.

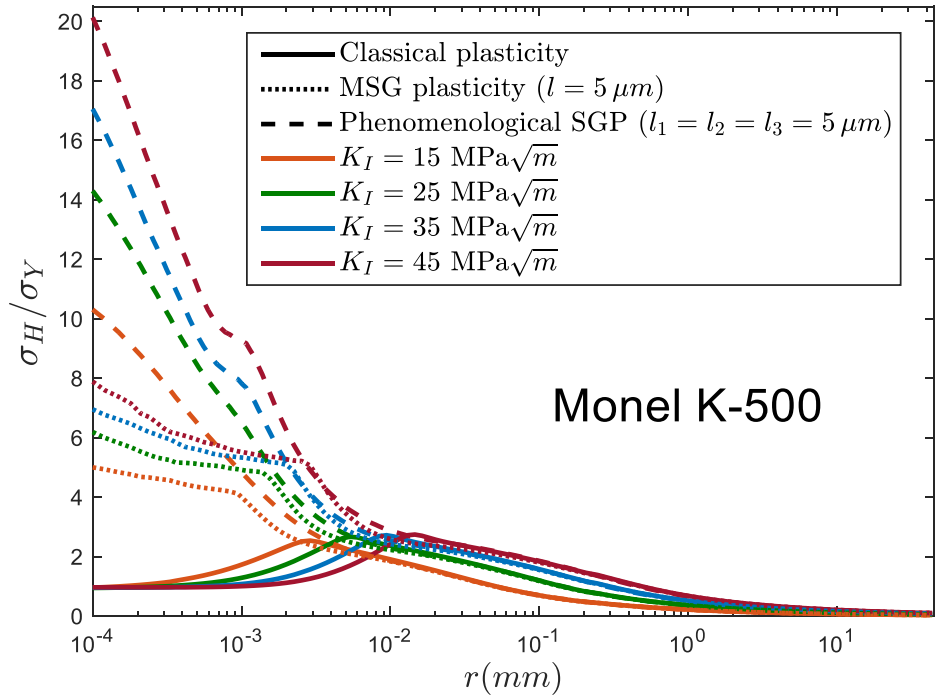

Figure 1. FEA calculated σ_H/σ_Y versus distance ahead of the blunted crack tip, $r$, for the range of K_I used in HEAC experiments with Monel K-500. Formulations include: MSGP ($l_{ref}$ = 5 μm), PSGP ($l_{ref} = l_1 = l_2 = l_3$ = 5 μm), and conventional plasticity. σ_Y in the flow rule for FEA [37] is equated to the measured tensile σ_YS, and the associated stress-strain relationship is essentially the same as the Ramberg-Osgood fit for Monel K-500.

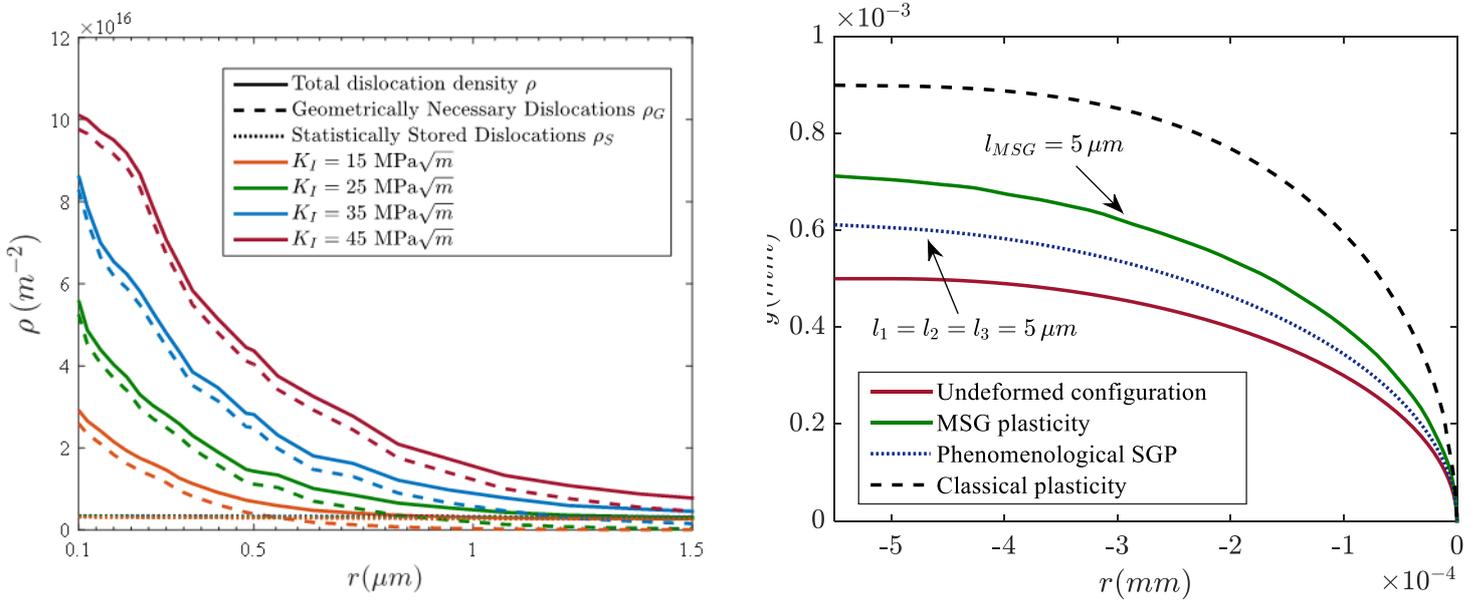

Figure 2. SGP-FEA calculations for Monel K-500 with $l_{ref}$ = 5 μm: (a) MSGP results showing ρ_S and ρ_G versus $r$ for the range of K_I used in the HEAC experiments, and (b) MSGP and PSGP predictions of blunt-crack opening shape for K_I = 15 MPa√m compared to the profile from classical plasticity.

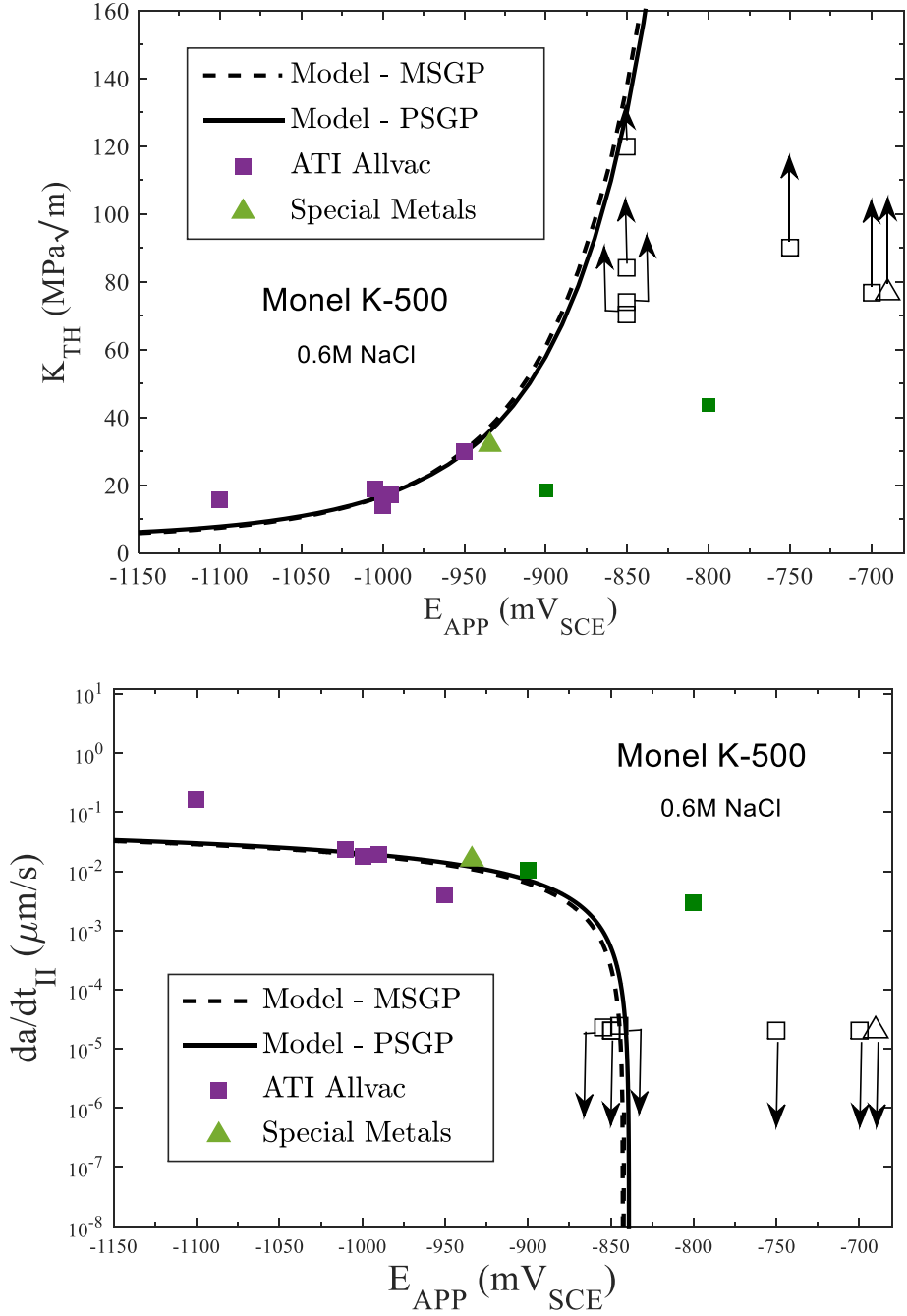

Figure 3. H-decohesion based predictions for Monel K-500 in 0.6M NaCl solution, calibrated by adjusting α in (1) to fit the average of replicate experimental measurements of $K_{TH}$ at $E_{APP}$ = -1.000 $V_{SCE}$ for $\sigma_H$ determined by PSGP (solid line, $\sigma_H = 8.15\sigma_Y$, $\alpha = 6.36$ MPa$\sqrt{m}$ (at frac H)$^{-1}$ and $C_{H\sigma-crit} = 407$ wppm), as well as MSGP (dashed line, $\sigma_H = 4.7\sigma_Y$, $\alpha = 37.59$ MPa$\sqrt{m}$ (at frac H)$^{-1}$ and $C_{H\sigma-crit} = 68$ wppm), each with $l_{ref}$ = 5 μm; (a) $K_{TH}$ versus $E_{APP}$, and (b) da/dt$_{II}$ versus $E_{APP}$. Other parameters are $k_{IG} = 0.880$ MPa$\sqrt{m}$ [45], $D_{H-EFF} = 1 \cdot 10^{-10}$ $cm^2/s$ [49], and $x_{crit} = 1$ μm [54].

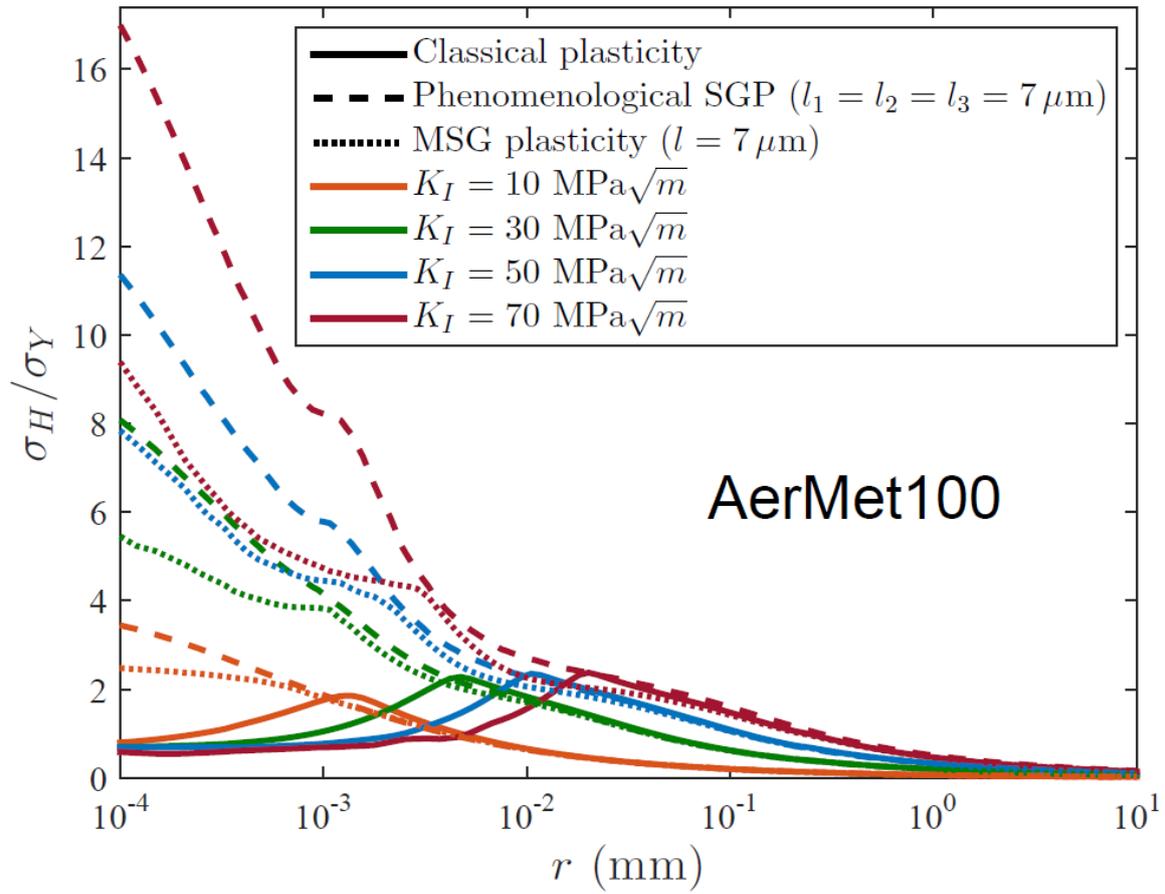

Figure 4. FEA calculated σ$_H$/σ$_Y$ versus distance ahead of the blunted crack tip, *r*, for the range of K$_I$ used in the HEAC experiments with AerMet100. Formulations include: MSGP ($l_{ref}$ = 7 μm), PSGP ($l_{ref}$ = $l_1$ = $l_2$ = $l_3$ = 7 μm) and conventional plasticity. The σ$_Y$ in the flow rule for FEA [37] is equated to measured tensile σ$_{YS}$ of 1725 MPa and the associated stress-strain relationship is essentially the same as the Ramberg-Osgood fit for AerMet™100.

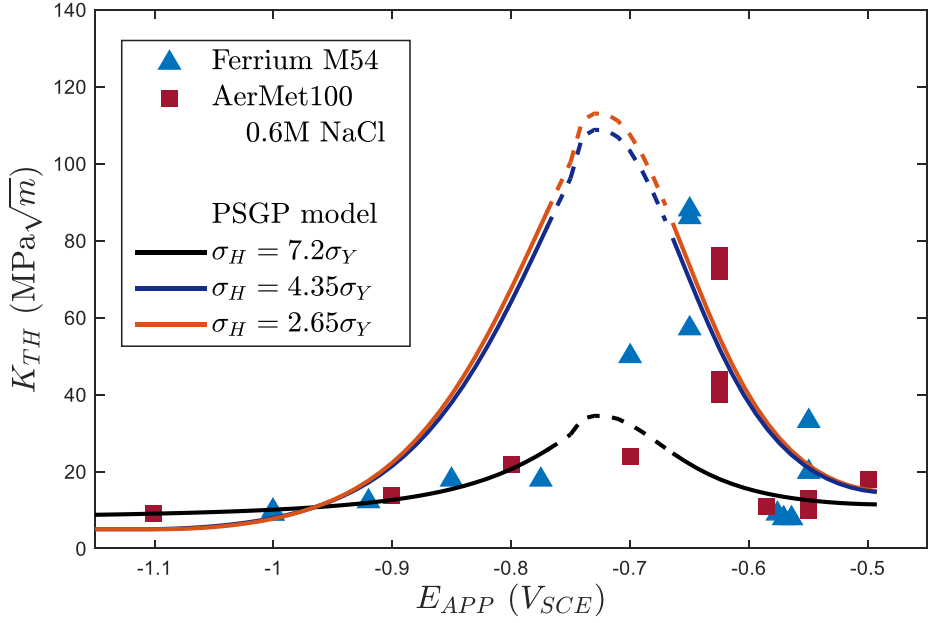

(a)

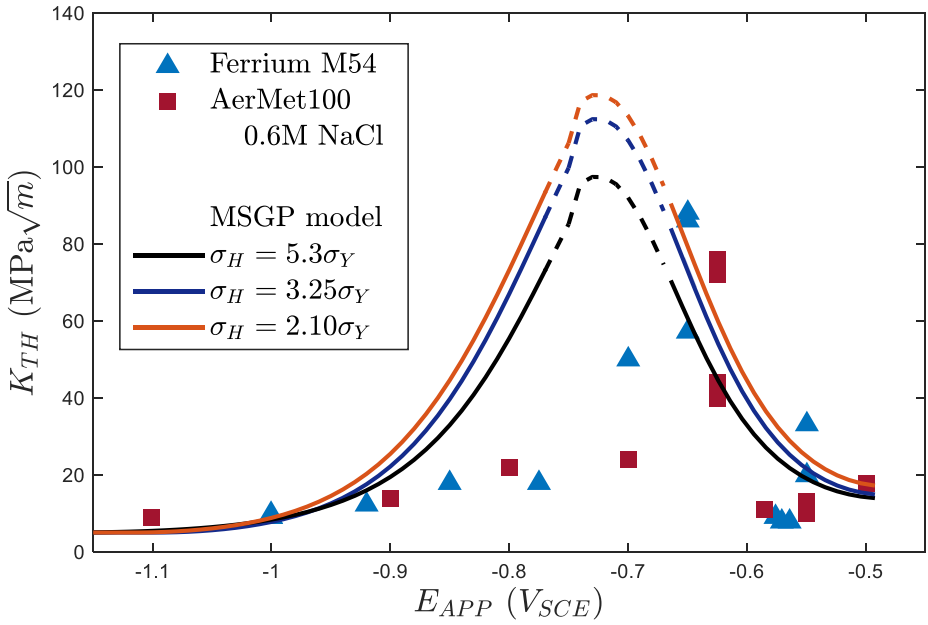

(b)

Figure 5. Predicted $K_{TH}$ versus $E_{APP}$ from (1) for AerMet™100 and Ferrium™M54 in 0.6M NaCl, calculated using upper bound $C_{H,Diff}$ from (6) and (7), and calibrated by averaging α by fitting to six $K_{TH}$ values measured at $E_{APP} \leq -0.9$ $V_{SCE}$; $k_{IG}$ = 1.145 MPa√m for each steel. The $σ_H$ is estimated from either: (a) PSGP or (b) MSGP FEA at K of 10 MPa√m (orange line: (a) $\bar{α}$ = 81.37 MPa√m (at frac H)$^{-1}$ and (b) $\bar{α}$ = 161.81 MPa√m (at frac H)$^{-1}$), 20 MPa√m (blue line: (a) $\bar{α}$ = 8.18 MPa√m (at frac H)$^{-1}$ and (b) $\bar{α}$ = 35.64 MPa√m (at frac H)$^{-1}$) and 40 MPa√m (black line: (a) $\bar{α}$ = 0.76 MPa√m (at frac H)$^{-1}$ and (b) $\bar{α}$ = 2.65 MPa√m (at frac H)$^{-1}$) . The $σ_H/σ_Y$ listed on each plot increased as $K_I$ rose from 10 to 20 to 40 MPa√m.

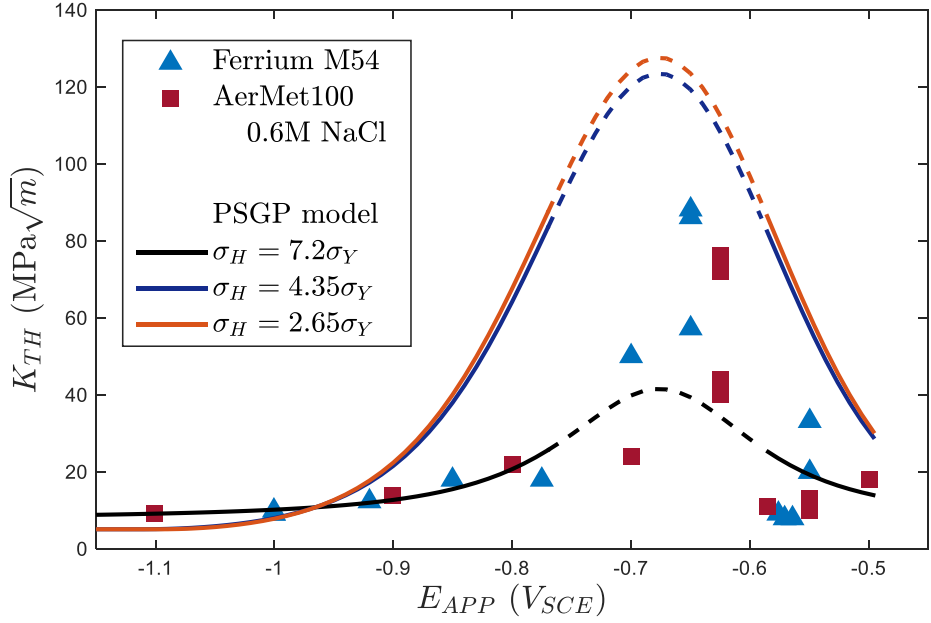

(a)

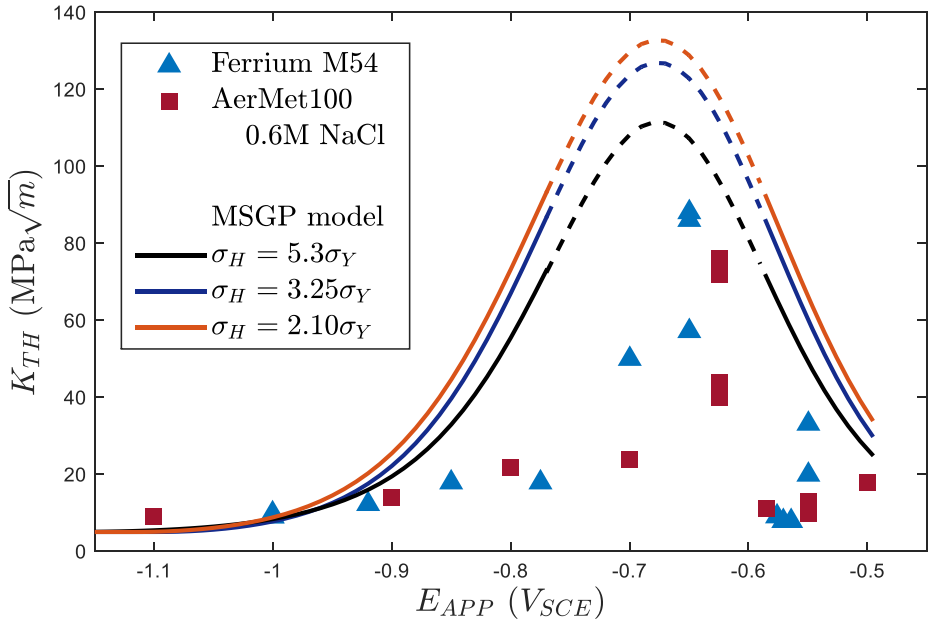

(b)

Figure 6. Predicted $K_{TH}$ versus $E_{APP}$ from (1) for AerMet™100 and Ferrium™M54 in 0.6M NaCl, calculated using lower bound $C_{H,Diff}$ from (6) and calibrated by averaging α from six experimental $K_{TH}$ values measured at $E_{APP} \leq$ -0.9 $V_{SCE}$; $k_{IG}$ = 1.145 MPa$\sqrt{m}$ for each steel. The $\sigma_H$ is estimated from either: (a) PSGP or (b) MSGP FEA at K of 10 MPa√m (orange line: (a) $\bar{\alpha}$ = 81.37 MPa$\sqrt{m}$ (at frac H)$^{-1}$ and (b) $\bar{\alpha}$ = 161.81 MPa$\sqrt{m}$ (at frac H)$^{-1}$), 20 MPa√m (blue line: (a) $\bar{\alpha}$ = 8.18 MPa$\sqrt{m}$ (at frac H)$^{-1}$ and (b) $\bar{\alpha}$ = 35.64 MPa$\sqrt{m}$ (at frac H)$^{-1}$) and 40 MPa√m (black line: (a) $\bar{\alpha}$ = 0.76 MPa$\sqrt{m}$ (at frac H)$^{-1}$ and (b) $\bar{\alpha}$ = 2.65 MPa$\sqrt{m}$ (at frac H)$^{-1}$) . The $\sigma_H/\sigma_Y$ listed on each plot increased as $K_I$ rose from 10 to 20 to 40 MPa√m.

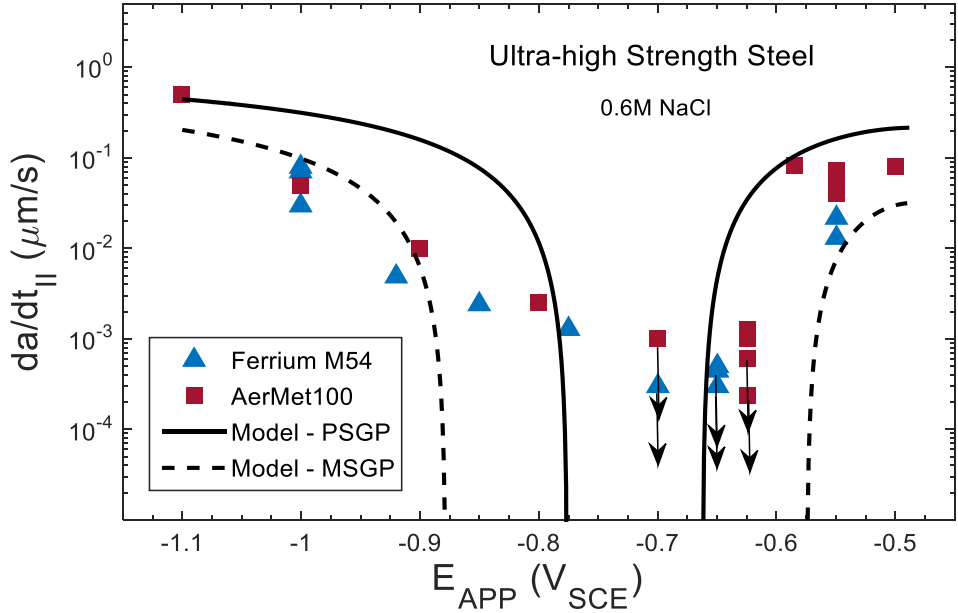

Figure 7. da/dt$_{II}$ versus E$_{APP}$ predicted from (2) with upper bound C$_{H,Diff}$ from (6) and (7) for AerMet$^{TM}$100 and Ferrium$^{TM}$M54 in 0.6M NaCl. The σ$_H$ is determined for K of 40 MPa√m using either PSGP (solid line, $C_{H\sigma-crit}$ = 18,867 wppm for σ$_H$/σ$_Y$ = 7.2) or MSGP (dashed line, $C_{H\sigma-crit}$ = 3,056 wppm for σ$_H$/σ$_Y$ = 5.3). Other parameters are $k_{IG}$ = 1.145 MPa√m, $D_{H-EFF}$ = 1·10$^{-9}$ $cm^2/s$ [51] and $x_{crit}$ = 1 $\mu m$ [54].

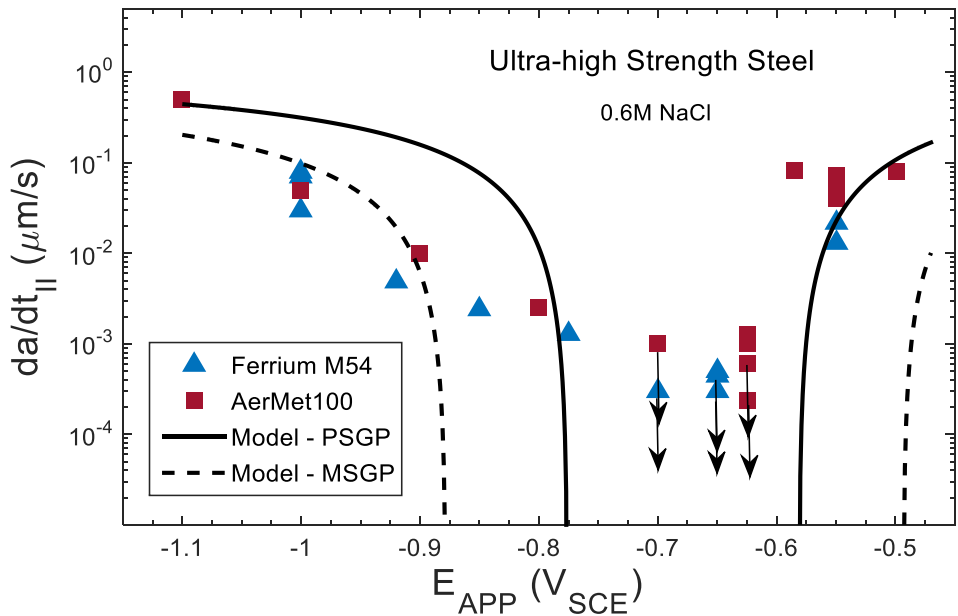

Figure 8. da/dt$_{II}$ versus E$_{APP}$ predicted from (2) with lower bound C$_{H,Diff}$ from (6) for AerMet$^{TM}$100 and Ferrium$^{TM}$M54 in 0.6M NaCl. The σ$_H$ is determined for K of 40 MPa√m using either PSGP (solid line, $C_{H\sigma-crit}$ = 18,867 wppm for σ$_H$/σ$_Y$ = 7.2) or MSGP (dashed line, $C_{H\sigma-crit}$ = 3,056 wppm for σ$_H$/σ$_Y$ = 5.3). Other parameters are $k_{IG}$ = 1.145 MPa√m, $D_{H-EFF}$ = 1·10$^{-9}$ $cm^2/s$ [51] and $x_{crit}$ = 1 $\mu m$ [54].